\documentstyle[12pt]{article}

\def\bseq{\begin{subequation}}  
\def\eseq{\end{subequation}}
\def\bsea{\begin{subeqnarray}}  
\def\esea{\end{subeqnarray}}

\def\beq{\begin{equation}}
\def\eeq{\end{equation}}
\def\eea{\end{eqnarray}}
\def\bq{\begin{quote}}
\def\eq{\end{quote}}

\newcommand{\EQ}{\begin{equation}}
\newcommand{\EN}{\end{equation}}
\newcommand{\bea}{\begin{eqnarray}}
\newcommand{\ena}{\end{eqnarray}}

\renewcommand{\a}{\alpha}
\renewcommand{\b}{\beta}

\renewcommand{\d}{\delta}
\newcommand{\th}{\theta}

\newcommand{\pa}{\partial}

\renewcommand{\l}{\lambda}

\newcommand{\m}{\mu}

\newcommand{\n}{\nu}

\newcommand{\p}{\pi}

\newcommand{\r}{\rho}

\newcommand{\s}{\sigma}

 \newcommand{ \Xb}{\bar{X}}
 \newcommand{\Db}{\bar{D}}
 
 \newcommand{\Psib}{\bar{\Psi}}

\newcommand{\pp}{\mid \!\!\! =}

\renewcommand{\thefootnote}{\fnsymbol{footnote}}

\begin{document}

\newpage
\begin{titlepage}
\begin{flushright}
{IFUM-464-FT}\\
{BRX-TH-356~~}\\
{hep-th/9403194}
\end{flushright}
\vspace{2cm}
\begin{center}
{\bf {\large  K\"{A}HLER POTENTIALS AND RENORMALIZATION GROUP FLOWS IN N=2
LANDAU-GINZBURG MODELS}}\\
\vspace{1.5cm}
M.T. GRISARU\footnote{
Work partially supported by the National Science Foundation under
grant PHY-92-22318.} \\
\vspace{1mm}
{\em Physics Department, Brandeis University, Waltham, MA 02254, USA}\\
\vspace{5mm}
and\\
\vspace{5mm}
D. ZANON\\
\vspace{1mm}
{\em  Dipartimento di Fisica dell'Universit$\grave{a}$ di Milano and INFN,
Sezione di Milano, Via Celoria 16, I-20133 Milano, Italy}
\vspace{1.1cm}\\
{{ABSTRACT}}
\end{center}

\bq
We examine the conditions for superconformal invariance and the specific form
of the K\"ahler potential for a  two-dimensional  lagrangian model with $N=2$
supersymmetry
and superpotential $gX^k$. Away from the superconformal point we study the
renormalization group flow  induced  by a particular class of K\"ahler
potentials. We find
trajectories which, in the infrared, reach the fixed point with a central
charge
whose value is that of the $N=2$, $A_{k-1}$ minimal model.
\eq

\vfill

\begin{flushleft}
March 1994

\end{flushleft}
\end{titlepage}

\newpage

\renewcommand{\thefootnote}{\arabic{footnote}}
\setcounter{footnote}{0}
\newpage
\pagenumbering{arabic}

In recent years the Landau-Ginzburg description of $N=2$ supersymmetric
theories
in two-dimensions
has been studied extensively
\cite{Kastor,Vafa,Howe,Cecotti,Marshakov,Liao,Witten}.
It is generally accepted that  these
models, with given superpotentials and suitable, though
unspecified K\"ahler potentials,  describe renormalization group
flows toward infrared fixed points which can be identified with $N=2$ minimal
models.  Along the flow trajectories the $N=2$ nonrenormalization theorem
ensures
that the form of the superpotentials remains unchanged while
the K\"ahler potentials adjust themselves in such a way that at
the fixed points the resulting actions describe superconformally
invariant systems. However, to the best of
our knowledge, no explicit lagrangian models have been constructed which
exhibit this behaviour. In this work we  present such a model. (In
a recent paper Fendley and Intrilligator
\cite{Fendley}  have studied N=2 flows in an exact S-matrix context.)

For simplicity we discuss primarily a system with a single chiral
superfield $X$, and the Landau-Ginzburg superpotential $gX^k$.  We first
examine the situation at the fixed point and show that
the condition for superconformal invariance determines the K\"ahler
potential up to an overall constant. In fact the
model is not conformal unless the
supercurrent is suitably improved and the construction of such an
improvement term is possible only if the K\"ahler potential
has a specific form.  (For models with more fields the condition
is less restrictive.) We then extend the analysis off-criticality and
consider a
lagrangian whose RG trajectories admit an IR fixed point where the
K\"ahler
potential takes the above-mentioned specific form.  At the IR
critical point the lagrangian is
the one used in \cite{Marshakov} (and also, in its Liouville version in
\cite{Liao}), but in contradistinction to these
references the normalization of the kinetic term is not arbitrary, but
fixed by the flow equations.

Our model has the conventional appearance
\EQ
{\cal S} = \int d^2x d^4 \th~ K(X, \Xb ) +\int d^2xd^2 \th~ W(X) +\int d^2x
d^2\bar{\th}~ \bar{W}(\Xb)
\EN
We use the following notation
\bea
x_{\pp} =\frac{1}{\sqrt{2}} (x_0+x_1) ~~~~,~~~~\pa_{\pp}= \frac{1}{\sqrt{2}}
(\pa_0+\pa_1) \nonumber\\
x_{=} =\frac{1}{\sqrt{2}} (x_0-x_1) ~~~~,~~~~\pa_{=}= \frac{1}{\sqrt{2}}
(\pa_0-\pa_1)
\ena
with
\EQ
\Box \equiv \pa^{\m}\pa_{\m}  =2\pa_{\pp}\pa_{=} ~~~~,~~~~ \pa_=
\frac{1}{x_{\pp} }= 2\pi i \d^{(2)}(x)
\EN
The superspace coordinates are $\th_+$, $\th_-$, $\bar{\th}_+$, $\bar{\th}_-$,
and the superspace covariant derivatives satisfy
\EQ
\{D_+, \bar{D}_+\} = i \pa_{\pp} ~~~~,~~~~\{D_-,\bar{D}_-\} = i \pa_=
\EN
with all other anticommutators vanishing. We also freely interchange
$\int d^2\th \Longleftrightarrow D_+D_- \equiv D^2$
and $\int d^2\bar{\th}  \Longleftrightarrow \bar{D}_+\bar{D}_-
\equiv \bar{D}^2$. Finally, for a kinetic term  $\int X \Xb $ the chiral
field propagator is
\EQ
<X(x, \th ) \Xb (0)> = -\frac{1}{2\pi} \bar{D}^2 D^2 \d^{(4)}(\th )
\ln[m^2(2x_{\pp}x_= +\ell ^2)]
\EN
where $m$ and $\ell$ are infrared and ultraviolet cutoffs respectively.

We find it convenient to discuss questions of superconformal invariance by
coupling the above
system to {\em linearized} $N=2$ supergravity. In doing so we include a
chiral  ``dilaton''
improvement term, so that the action takes the form
\bea
{\cal S}&=& \int d^2x d^4 \th E^{-1} K[(1+iH.\pa )X, (1-iH.\pa )\Xb ] +\int
d^2x d^2 \th e^{-2 \s}
W(X) \nonumber\\
&+&\int d^2x d^2 \bar{\th} e^{-2\bar{\s}} \bar{W}(\Xb )
+\int d^2x d^2\th~ R~ \Psi(X) +\int d^2x d^2 \bar{\th}~\bar{R}~\bar{\Psi}(\Xb)
\ena
where, at the linearized level,
\bea
E^{-1}&=&1-[\bar{D}_+,D_+]H_=-[\bar{D}_-,D_-]H_{\pp} \nonumber\\
R&=& 4\bar{D}_+\bar{D}_- [\bar{\s} +D_+\bar{D}_+H_=+D_-\bar{D}_-H_{\pp}]
\nonumber\\
\bar{R}&=&4D_+D_-[ \s  -\bar{D}_+D_+H_=  -\bar{D}_-D_-H_{\pp}]
\ena
Here $H$ is the supergravity  potential, while $\s$ is the (chiral) compensator
\cite{us}. \footnote{In ref. \cite{us}  we were working in conformal gauge so
that only the compensator $\s$
was present, but it is easy to include the $H$ field by solving the
constraints at the linearized
level; see also ref. \cite{Aziz}. }

The superconformal properties of the system are encoded in the supercurrent
\EQ
J_{\pp} \equiv \frac{\d {\cal S}}{\d H_=} =2 [D_+X \Db_+\Xb K_{X\Xb} -2
\Db_+D_+\Psi +2D_+\Db_+\Psib]
\EN
and the associated supertrace
\EQ
J \equiv \frac{\d {\cal S}}{\d \s}= -2[W -2 \Db_+\Db_-\Psib]
\EN
We have introduced the K\"ahler metric
\EQ
K_{X\Xb}= \frac{\pa^2K}{\pa X \pa \Xb}
\EN

Superconformal invariance requires the supertrace $J$ to vanish. For the
superpotential $W=gX^k$ the equations of motion (with the notation $K_X =
\pa_XK$, etc.)
\EQ
\Db_+\Db_- K_X +W_X=0
\EN
give $W = - \frac{1}{k} \Db_+ \Db_- (X K_X)$. The condition $J=0$ implies then
\EQ
X K_X = -2k\Psib (\Xb )
\EN
modulo a {\em linear} superfield (annihilated by $\Db_+ \Db_-$) which gives
no contributions to
the action. We have assumed that $\Psib$ is local, and (anti)chirality and
dimensionality require it to be just a function of $\Xb$.
Integrating with respect to $X$ and imposing also $\bar{J}=0$ we find
\bea
K &=& \a \ln X \ln \Xb \nonumber\\
\Psib &=& - \frac{\a}{2k} \ln \Xb
\ena
with arbitrary constant $\a$. (With this solution the improvement terms in
 the supercurrent $J_{\pp}$ can be rewritten as  $\frac{2}{k}
[D_+(X\bar{D}_+K_X
)-\bar{D}_+(\Xb D_+K_{\Xb})]$. ) Using the field redefinition $X \equiv
 e^{\Phi}$ the
corresponding lagrangian
can be recast in Liouville form.

We note that  for such a K\"ahler potential
conformal invariance is not broken by quantum corrections since
the one-loop $\b$-function, proportional to the Ricci tensor $R_{X\Xb}=
\pa_X\pa_{\Xb} tr \ln K_{X\Xb} $,
vanishes and all the higher-loop contributions which
involve the Riemann tensor, trivially vanish as well.
Moreover, while in the bosonic or in the $N=1$ supersymmetric theories
the dilaton
term contributes to the metric $\b$-function, in the $N=2$ case
no metric-dilaton mixing occurs due to the chirality of
$\Psi$.

We describe briefly the situation for a model with two fields.
For example, in the case of the superpotential
\EQ
W= X^{k+1} +XY^2
\EN
the construction of an improvement term is possible only if the K\"ahler
potential satisfies
\EQ
XK_X+ \frac{k}{2}YK_Y=-2(k+1) \Psib (\Xb , \bar{Y})
\EN
As a partial differential equation in $X$ and $Y$ this equation has many
solutions, but
these are severely restricted by requiring that the resulting metric be
Ricci-flat. One
finds in general the K\"ahler potential (assumed to be symmetric in chiral and
antichiral
fields)
\EQ
K(X,Y, \Xb ,\bar{Y}) = \frac{A}{\n^2} \left( \ln \frac{X^{\frac{k}{2}}}{Y} \ln
\frac{\Xb^{\frac{k}{2}}}{\bar{Y}}\right)^{\n}
+ B\ln XY \ln \Xb \bar{Y}
\EN
where  $A \neq 0$, $B\neq 0$,  and $\n$ are arbitrary constants.  (For $\n =0$
the first term is replaced by the square of the  logarithm of the expression
in parantheses.)
For $\n=1$ the  field redefinitions
\EQ
\frac{X^{\frac{k}{2}}}{Y} \equiv e^{\Phi_1} ~~~~~,~~~~~ XY \equiv e^{\Phi_2}
\EN
recast the  lagrangian, including the superpotential, in
Toda field theory form.

\vspace{2.0cm}

We consider now a model which flows in the IR region to
the superconformal theory defined above. It is described by the superpotential
$gX^k$
and the K\"ahler metric
\EQ
K_{X\Xb} = \frac{1}{1+bX\Xb +c (X\Xb)^2}
\EN
corresponding to the K\"ahler potential
\EQ
K = \int dX d\Xb K_{X\Xb}=
 X\Xb  -\frac{b}{4}(X \Xb )^2 +\frac{b^2-c}{9}
(X\Xb )^3+\cdots
\EN

The divergences of the model require renormalization of the parameters $b$,
$c$, and
 wave-function renormalization. However, it is convenient to rescale the
field, $X \rightarrow  a^{-\frac{1}{2}}X$, so that the K\"ahler metric and
superpotential become
(with a redefinition of the parameter $c$)
\EQ
K_{X\Xb} = \frac{1}{a+bX\Xb +c (X\Xb)^2} ~~~~~,~~~~~ g a^{-\frac{k}{2}} X^k
\EN
(A related metric, with $a=c$,
has been discussed  in a bosonic $\s$-model
context by
Fateev {\em et al} \cite{Fateev}. The authors of ref. \cite{Fendley} have
speculated on the
relevance of such metrics for studying N=2 flows.).

The model is rendered finite in $\s$-model fashion by renormalizing the
metric
 including the parameter $a$  (this is equivalent to  wave-function
renormalization)
and therefore, because the superpotential is not renormalized, the coupling
constant $g$.
At the one-loop level one finds the  divergent contribution, proportional to
the Ricci tensor,

\EQ
-(\frac{1}{2\pi} \ln m^2 \ell^2)R_{X\Xb}=(\frac{1}{2\pi} \ln m^2 \ell^2)
\frac{ab+4acX\Xb+bc(X\Xb)^2}{[a+bX\Xb+c(X\Xb)^2]^2}
\EN
This divergence
can be cancelled by  expressing the original parameters in the
classical lagrangian
in terms of renormalized ones
\bea
a&=&Z_a a_R~~~~,~~~~b=Z_b b_R~~~~,~~~~c=Z_c c_R \nonumber\\
g&=&\m Z_g g_R
\ena
with $\m$ the mass scale, and $Z_gZ_a^{-\frac{k}{2}}=1$
as required by the $N=2$ nonrenormalization theorem.
The renormalization constants are
\bea
Z_a&=& 1+ b(\frac{1}{2\pi} \ln \m^2 \ell^2)\nonumber\\
Z_b&=& 1+\frac{4ac}{b} (\frac{1}{2\pi} \ln \m^2 \ell^2)\nonumber\\
Z_c&=& 1+b(\frac{1}{2\pi} \ln \m^2 \ell^2)\nonumber\\
Z_g&=& 1+\frac{bk}{2}(\frac{1}{2\p} \ln \m^2 \ell^2)
\ena
In the following we shall drop the subscript on the renormalized parameters.


Defining $t=\ln \m$, the renormalized
parameters satisfy the following renormalization group equations
\bea
\frac{da}{dt}&=& -\frac{1}{\pi}ab \nonumber\\
\frac{db}{dt}&=& -\frac{4}{\pi}ac \nonumber\\
\frac{dc}{dt}&=& -\frac{1}{\pi}cb \nonumber\\
\frac{dg}{dt}&=&-(1+\frac{b}{2\p}k)g
\ena
These equations  have two invariants, the ratio
\EQ
\frac{a}{c}=\r
\EN
and the combination, which we choose to make positive and parametrize suitably,
\EQ
b^2-4ac = b^2-4\r c^2 =(\pi \l)^2
\EN
In the $b$-$c$ plane we obtain two types of trajectories, hyperbolas or
ellipses, depending on the
sign of $\r$, as depicted in Fig. 1.

\let\picnaturalsize=N
\def\picsize{4.0in}
\def\picfilename{flow1.eps}
\ifx\nopictures Y\else{\ifx\epsfloaded Y\else\input epsf \fi
\let\epsfloaded=Y
\centerline{\ifx\picnaturalsize N\epsfxsize \picsize\fi
\epsfbox{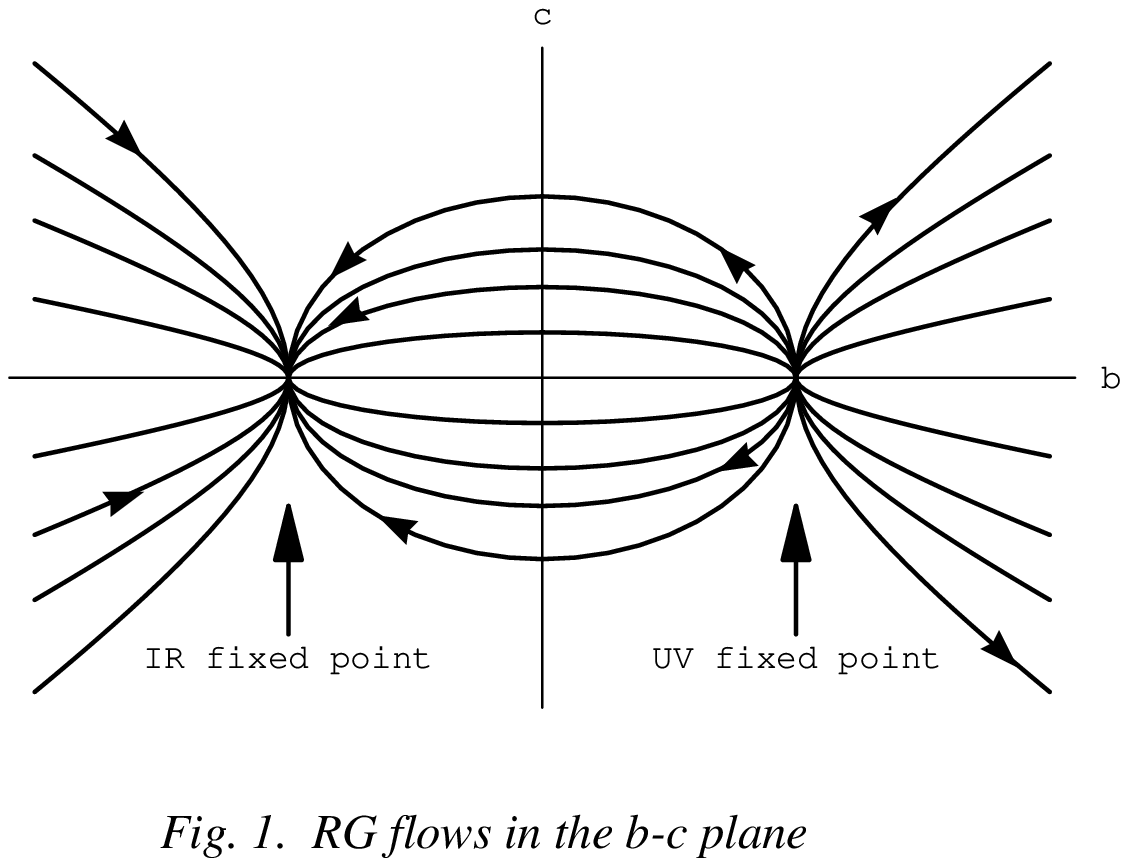}}}\fi

Since we are interested in trajectories
with two fixed points we write  the elliptical solutions,
with $\r <0$.
(The  bosonic model studied in \cite{Fateev},  written in a different
coordinate
system,  has $\r=1$.)
\bea
b(t) &=& \pi \l  \tanh \l t \nonumber\\
a(t) &=& \pm \frac{\pi \l  \sqrt{-\r}}{2} (\cosh \l t)^{-1} \nonumber \\
c(t) &=& \mp  \frac{\pi \l}{2 \sqrt{-\r}} (\cosh \l t)^{-1}\nonumber\\
g(t) &=& g_0 e^{-t}[\cosh \l t ]^{-\frac{k}{2}}
\ena

Conformal invariance is achieved at the zeroes of the coupling
$\b$-functions in eq.(24). In particular
we are looking for trajectories  which lead to a nontrivial
IR fixed point for the
coupling constant $g$, i.e. such that $b(t) \rightarrow -\frac{2\pi}{k}$ as
$t \rightarrow - \infty$. We achieve this by choosing
\EQ
\l = \frac{2}{k}
\EN
In this case the superfield $a^{-\frac{1}{2}} X$
acquires anomalous dimension $1/k$ in the
corresponding IR conformal theory, while $a$ and $c$ flow to zero.
Therefore, the effective lagrangian with K\"ahler potential
$K(X,\Xb,a(t), b(t), c(t))$
and superpotential $W(X,g(t),a(t))$ has the following behaviour
 in the infrared,
\EQ
t \rightarrow - \infty~~~~,~~~~
K(t) \rightarrow  -\frac{k}{2\pi} \ln X \ln \Xb  ~~~~,~~~~ W(t)\rightarrow
g_0 X^k
\EN
The improvement term at the IR fixed point has $\Psi = \frac{1}{4\pi} \ln X$.

Changing variables, $X \equiv e^{\Phi}$, leads to the  Liouville lagrangian
\EQ
{\cal L} = -\frac{k}{2\pi} \bar{\Phi}\Phi +g_0 e^{k \Phi}
\EN
with negative kinetic term and {\em with
normalization determined
by the superpotential} (cf. \cite{Marshakov,Liao}).

We emphasize  that imposing conformal invariance at the one-loop level, i.e.
$R_{X\Xb}=0$, is sufficient to insure the absence of divergences at higher-loop
orders. Thus at the conformal point we obtain exact, all-order results.
(In these models there are no ``nonperturbative'' divergences.)

We briefly describe a generalization which allows us  to discuss the
stability of the IR fixed point. We consider a model
 with K\"ahler metric
\EQ
K_{X\Xb} = \frac{1}{\sum a_n (X\Xb )^n}
\EN
It is easy to verify that if the sum in the denominator is finite (but contains
more than the first three terms considered in  eq. (20)) the model is not
renormalizable by a redefinition of the parameters. If the sum is infinite one
computes the one-loop divergence proportional to the Ricci
tensor and after renormalization one is led to the flow equations
\bea
\frac{dg}{dt} &=& -(1+\frac{a_1}{2\pi}k)g \nonumber\\
\frac{da_0}{dt} &=& -\frac{1}{\pi} a_0a_1 \nonumber\\
\frac{da_1}{dt} &=& -\frac{4}{\pi}a_0a_2 \nonumber\\
\frac{da_2}{dt} &=& -\frac{1}{\pi}(a_1a_2+9a_0a_3) \nonumber\\
\frac{da_3}{dt} &=& -\frac{1}{\pi} (4a_1a_3+16a_0a_4) \nonumber\\
\frac{da_4}{dt} &=& -\frac{1}{\pi}(3a_2a_3 +9a_1a_4+25a_0a_5)
\ena
etc.
(In the last equation  the presence of the term $a_2a_3$ makes it obvious why
the model
is not renormalizable if one considers a finite sum with more than the original
three terms. )

If we want a nontrivial IR fixed point for the coupling constant
$g$, all the $a_n$ must flow to zero except for $a_1= -\frac{2\pi}{k}$, so that
we recover the
special case
studied above. Furthermore, by linearizing the flow equations around these
values
 it is possible to determine that
this fixed point is stable under perturbations with  the $a_n \neq 0$.

We compute now the central charge of our original model, from the coupling
to supergravity. We are looking for contributions to the supergravity
effective action of the form $R
\Box ^{-1} \bar{R}$.  They can be determined by
contributions to the $H_=$ self-energy, from which the covariant expression can
be
reconstructed. For one-loop contributions  the relevant vertices
are obtained from the coupling to $H_=$ in the
K\"ahler potential  while tree-level contributions  come from the direct
dilaton coupling in the improvement term.

We compute away from the fixed point, using an effective
configuration-space propagator
\EQ
<X(x, \th ) \Xb (x', \th ')> = -\frac{K^{X\Xb}}{2\pi} \bar{D}^2 D^2
\d^{(4)}(\th  - \th ') \ln\{m^2[2(x-x')_{\pp} (x-x')_= +\ell ^2]\}
\EN
where $K^{X\Xb}$ is the inverse of the K\"ahler metric
(cf \cite{4loop} eq. (3.13);
additional  terms, involving derivatives of the K$\ddot{\rm a}$hler metric
in the propagator do not
give relevant contributions).  The couplings to $H_=$ can be read from the
action in eq. (6) or,
equivalently, from the supercurrent. From  the K\"ahler potential
we have
\EQ
2i  \int d^4 \th ~ H_= D_+X \Db_+\Xb K_{X\Xb}
\EN
giving the one-loop contribution
\EQ
-\frac{1}{\pi^2} H_=\frac{\Db_- D_-\Db_+ D_+}{(x-x')^2_{\pp} }H_=
\EN
The coupling from the dilaton term is
\EQ
4 \int d^4 \th (\Psib  - \Psi )\pa_{\pp}H_=
\EN
and
leads to the contribution
\EQ
-\frac{16}{\pi} \Psi_X K^{X\Xb}\Psib_{\Xb} H_=\frac{\Db_- D_-\Db_+ D_+}
{(x-x')^2_{\pp} }H_=
\EN

Using the relation between the K$\ddot{\rm a}$hler potential and the
improvement term at the fixed point we obtain then the total contribution
\EQ
-\frac{1}{\pi^2}H_=\frac{\Db_- D_-\Db_+D_+}{(x-x')^2_{\pp}}
H_=\left(1-\frac{2}{k}\right)\Rightarrow
\frac{1}{4\pi}  R \frac{1}{\Box} \bar{R} \left(1-\frac{2}{k}\right)
\EN
and the correct central charge for the $N=2$, $A_{k-1}$ minimal model,
\EQ
c=1-\frac{2}{k}
\EN

Returning to the flows in eq. (27)  we note that moving
away from the IR fixed point along the RG trajectory toward the UV region, we
reach a value of  $t$, namely
\EQ
t= \frac{k}{2} \ln \frac{ X \Xb}{\sqrt{-\r}}
\EN
 where
the effective K\"ahler metric becomes singular. Therefore even if the
trajectories
we have been considering display an UV fixed point we cannot actually reach it.
Had we started past the singularity, in the UV region, we could follow the flow
to the UV fixed point, $t \rightarrow + \infty$ and find
\EQ
K(t) \rightarrow  \frac{k}{2\pi} \ln X \ln \Xb  ~~~~,~~~~ W(t)\rightarrow
g_0 X^k
\EN
It is interesting to observe that
the  central charge for this conformal field theory,
\EQ
c=1+\frac{2}{k} ~,
\EN
corresponds to the value of $c$ for the analytic continuation of
the N=2 minimal models.

\vspace{1.5cm}

{\bf Acknowledgment} D. Zanon thanks the Physics Department of
Harvard University for
hospitality during the period when some of this work was done. We thank
M. Bershadsky, M. Raciti and C. Vafa for useful  discussions.

\end{document}